\documentclass[11pt]{article}
\usepackage{amsmath,amssymb}
\usepackage[dvips]{epsfig}
\usepackage{graphicx}

\newtheorem{theorem}{Theorem}

\newtheorem{remark}{Remark}

{\hspace*{\fill}$\rule{.3\baselineskip}{.35\baselineskip}$\end{trivlist}}

\newcommand{\R}{\mathbb{R}}

\renewcommand{\geq}{\geqslant}
\renewcommand{\leq}{\leqslant}

\renewcommand{\phi}{\varphi}
\newcommand{\be}{\begin{eqnarray}}
\newcommand{\ee}{\end{eqnarray}}

\newcommand{\eps}{\varepsilon}

\begin{document}

\title{\bf Asymptotic properties of excited states \\ in the Thomas--Fermi limit}

\author{Dmitry Pelinovsky \\
{\small Department of Mathematics, McMaster
University, Hamilton, Ontario, Canada, L8S 4K1}  }

\date{\today}
\maketitle

\begin{abstract}
Excited states are stationary localized solutions of the
Gross--Pitaevskii equation with a harmonic potential and a
repulsive nonlinear term that have zeros on a real axis. Existence
and asymptotic properties of excited states are considered in the
semi-classical (Thomas-Fermi) limit. Using the method of
Lyapunov--Schmidt reductions and the known properties of the
ground state in the Thomas--Fermi limit, we show that excited
states can be approximated by a product of dark solitons
(localized waves of the defocusing nonlinear Schr\"{o}dinger
equation with nonzero boundary conditions) and the ground state.
The dark solitons are centered at the equilibrium points where a
balance between the actions of the harmonic potential and
the tail-to-tail interaction potential is achieved.
\end{abstract}

\section{Introduction}

The defocusing nonlinear Schr\"{o}dinger equation is derived in
the mean-field approximation to model Bose--Einstein condensates
with repulsive inter-atomic interactions between atoms. This
equation is referred in this context to as the Gross--Pitaevskii
equation \cite{PitStr}. When the Bose--Einstein condensate is
trapped by a magnetic field, the Gross--Pitaevskii equation has a
harmonic potential. In the strongly nonlinear limit, referred to
as the Thomas--Fermi limit \cite{Fermi,Thomas}, the Bose--Einstein
condensate is a nearly compact cloud, which may contain localized
dips of the atomic density. The nearly compact cloud is modeled by
the ground state of the Gross--Pitaevskii equation, whereas the
localized dips are modeled by the excited states. Asymptotic
properties of the stationary excited states in the Thomas--Fermi
limit are analyzed in this article.

The Gross--Pitaevskii equation with a harmonic potential and a
repulsive nonlinear term can be rewritten in the form
\begin{equation}
\label{GP} i \eps u_t + \eps^2 u_{xx} + (1 - x^2 - |u|^2) u = 0,
\end{equation}
where $\eps > 0$ is a small parameter to model the Thomas--Fermi
asymptotic regime. Let $\eta_{\eps}$ be the real positive solution
of the stationary equation
\begin{equation}
\label{stationaryGP} \eps^2 \eta_\eps''(x) + (1- x^2 -
\eta_\eps^2(x)) \eta_{\eps}(x) =0, \quad x \in \mathbb{R}.
\end{equation}
Main results of Ignat \& Millot \cite{IM,IM2} and Gallo \&
Pelinovsky \cite{GalPel2} state that for any sufficiently small $\eps > 0$
there exists a unique smooth positive solution $\eta_{\eps} \in {\cal
C}^{\infty}(\mathbb{R})$ that decays to zero as $|x| \to \infty$
faster than any exponential function. The ground state
converges pointwise as $\eps \to 0$ to the compact Thomas--Fermi
cloud
\begin{equation}
\label{Thomas-Fermi}
\eta_0(x) := \lim_{\eps \to 0} \eta_{\eps}(x) = \left\{ \begin{array}{cl} (1 - x^2)^{1/2}, \;\; & \mbox{for} \;\; |x| < 1, \\
0, \;\; & \mbox{for} \;\; |x| > 1. \end{array} \right.
\end{equation}
The ground state and the convergence of $\eta_{\eps}$ to $\eta_0$ is characterized by the following properties:

\begin{itemize}
\item[P1] $0 < \eta_{\eps}(x) \leq 1$ for any $x \in \mathbb{R}$.

\item[P2] For any small $\eps > 0$ and any compact subset $K
\subset (-1,1)$, there is $C_K > 0$ such that
\begin{equation}
\label{C-K-bound} \| \eta_{\eps} - \eta_0 \|_{C^1(K)} \leq C_K
\eps^2.
\end{equation}

\item[P3] For any small $\eps > 0$, there is $C > 0$ such that
\begin{equation}
\label{L-infty-bound} \| \eta_{\eps} - \eta_0 \|_{L^{\infty}} \leq
C \eps^{1/3}, \quad \| \eta_{\eps}' \|_{L^{\infty}} \leq C
\eps^{-1/3}.
\end{equation}

\item[P4] There is $C > 0$ such that $\eta_{\eps}(x) \geq C \eps^{1/3}$ for any $|x| \leq 1 + \eps^{2/3}$.
\end{itemize}

Properties [P1] and [P2] follow from Proposition 2.1 in \cite{IM}.
Properties [P3] and [P4] follow from Theorem 1 in \cite{GalPel2}.
To clarify the proof of bound (\ref{L-infty-bound}), we represent
the ground state $\eta_{\eps}(x)$ in the equivalent form
\begin{equation}
\label{representation-ground}
\eta_{\eps}(x) = \eps^{1/3} \nu_{\eps}(y), \quad y =
\frac{1-x^2}{\eps^{2/3}},
\end{equation}
where $\nu_{\eps}(y)$ solves
$$
4 (1 - \eps^{2/3} y) \nu_{\eps}''(y) - 2 \eps^{2/3} \nu_{\eps}'(y)
+ y \nu_{\eps}(y) - \nu_{\eps}^3(y) = 0, \quad
y \in (-\infty,\eps^{-2/3}).
$$
Let $\nu_0(y)$ be the unique solution of the Painlev\'e--II equation
$$
4 \nu_0''(y) + y \nu_0(y) - \nu_0^3(y) = 0, \quad y \in \R,
$$
such that $\nu_0(y) = y^{1/2} + {\cal O}(y^{-1})$ as $y \to \infty$ and $\nu_0(y)$
decays to zero as $y \to -\infty$ faster than any exponential function.
By Theorem 1 in \cite{GalPel2}, $\nu_{\eps}$ is a ${\cal C}^{\infty}$ function on
$(-\infty,\eps^{-2/3}]$, which is expanded into the asymptotic series for
any fixed $N \geq 0$:
\begin{equation}
\label{asymptotic-expansion-nu} \nu_{\eps}(y) = \sum_{n = 0}^N
\eps^{2n/3} \nu_n(y) + \eps^{(2N+1)/3} R_{N,\eps}(y),
\end{equation}
where $\{ \nu_n \}_{n = 1}^N$ are uniquely defined
$\eps$-independent $C^{\infty}$ functions on $\R$  and
$R_{N,\eps}(y)$ is the remainder term on $(-\infty,\eps^{-2/3}]$.
It was proved in \cite{GalPel2} that $U_{N,\eps}(z) =
R_{N,\eps}(\eps^{-2/3}- \eps^{2/3} z^2)$ is uniformly bounded for
small $\eps > 0$ in $H^2(\R)$-norm. If we denote $u_{N,\eps}(x) =
U_{N,\eps}(\eps^{-2/3}x) = R_{N,\eps}(y)$, then the above arguments shows that
there is $C_N > 0$ such that
$$
\| u_{N,\eps} \|_{L^{\infty}} \leq C_N, \quad \| u_{N,\eps}'
\|_{L^{\infty}} \leq C_N \eps^{-2/3}.
$$
For any fixed $N \geq 0$, it follows from the above bounds that
the remainder term $\eps^{(2N+1)/3} u_{N,\eps}(x)$ is smaller in ${\cal C}^1(\R)$
norm than the leading-order term $u_0(x) = \nu_0(\eps^{-2/3} -
\eps^{-2/3} x^2)$. The error estimate (\ref{L-infty-bound})
follows from (\ref{representation-ground}),
(\ref{asymptotic-expansion-nu}), and the fact that $\sup_{y \in
\R^+} | \nu_0(y) - y^{1/2} | < \infty$.

We shall consider excited states of the Gross--Pitaevskii equation
(\ref{GP}), which are real non-positive solutions of the stationary
equation
\begin{equation}
\label{stationaryGPexc} \eps^2 u_\eps''(x) + (1- x^2 -
u_\eps^2(x)) u_{\eps}(x) =0, \quad x \in \mathbb{R}.
\end{equation}
We classify the excited states by the number $m$ of zeros of
$u_{\eps}(x)$ on $\R$. A unique solution with $m$ zeros exists
near $\eps = \eps_m$ for $\eps < \eps_m$ by the local bifurcation
theory \cite{Kurth}, where $\eps_m$ is computed from the linear
theory as $\eps_m = \frac{1}{1 + 2 m}$, $m \in \mathbb{N}$.
Because of the symmetry of the harmonic potential, the $m$-th
excited state is even on $\mathbb{R}$ for even $m \in \mathbb{N}$
and odd on $\mathbb{R}$ for odd $m \in \mathbb{N}$.

This paper continues the previous research on the ground state in
the Thomas--Fermi limit that was developed by Gallo \& Pelinovsky
in \cite{GalPel1,GalPel2}. We focus now on the existence and
asymptotic properties of the excited states as $\eps \to 0$. Using
the method of Lyapunov--Schmidt reductions, we show that the
$m$-th excited state is approximated by a product
of $m$ dark solitons (localized waves of the defocusing nonlinear
Schr\"{o}dinger equation with nonzero boundary conditions) and the
ground state $\eta_{\eps}$. The dark solitons are centered at the
equilibrium points where a balance between the actions of the
harmonic potential and the tail-to-tail interaction potential
is achieved.

Note that this paper gives a rigorous justification of the
variational approximations found by Coles {\em et al.} in
\cite{CKP}, where the $m$-th excited states was approximated by
a variational ansatz in the form of a product of $m$ dark
solitons with time-dependent parameters and the ground state.
Time-evolution equations for the parameters of the variational
ansatz were found from the Euler-Lagrange equations. Critical
points of these equations give approximations of the equilibrium
positions of the dark solitons relative to the center of the
harmonic potential and to each others, whereas the linearization
around the critical points give the frequencies of oscillations of
dark solitons near such equilibrium positions. Variational approximations were found
in \cite{CKP} to be in excellent agreement with numerical solutions
of the stationary equation (\ref{stationaryGPexc}).

This article is organized as follows. The first excited state
centered at $x = 0$ is considered in Section 2. Although existence
of this solution can be established from the calculus of
variations, we develop the fixed-point iteration scheme to study
this solution as $\eps \to 0$. The second excited state is
approximated in Section 3. We will work with the method of
Lyapunov--Schmidt reductions to find the equilibrium position of
two dark solitons as $\eps \to 0$. Section 4 discusses the
existence results for the general $m$-th excited state
with $m \geq 2$.

Before we proceed with main results, let us discuss some
notations. If $A$ and $B$ are two quantities depending on a
parameter $\eps$ in a set $\mathcal{E}$, the notation
$A(\eps)=\mathcal{O}(B(\eps))$ as $\eps\to 0$ indicates that
$A(\eps)/B(\eps)$ remains bounded as $\eps\to 0$. If $A(x,\eps)$
depends on $x \in \R$ and $\eps \in \mathcal{E}$, the notation
$A(\cdot,\eps)=\mathcal{O}_{L^{\infty}}(B(\eps))$ as $\eps\to 0$
indicates that $\| A(\cdot,\eps) \|_{L^{\infty}} /B(\eps)$ remains
bounded as $\eps\to 0$. Different constants are denoted with the
same symbol $C$ if they can be chosen independently of the small
parameter $\eps$.

\section{First excited state}

The first excited state is an odd solution of the stationary equation (\ref{stationaryGPexc}) such that
\begin{equation}
\label{properties-u} u_{\eps}(0) = 0, \quad u_{\eps}(x) > 0 \;\; \mbox{\rm for all} \;\; x > 0, \quad \mbox{\rm
and} \quad \lim_{x \to \infty} u_{\eps}(x) = 0.
\end{equation}
Variational theory can be used to prove existence of this
solution, similar to the analysis of Ignat \& Millot in \cite{IM2}.
Since we are interested in asymptotic properties of the first
excited state as $\eps \to 0$, we will obtain both existence and
convergence results from the fixed-point arguments. Our main
result is the following theorem.

\begin{theorem}
For sufficiently small $\eps > 0$, there exists a unique solution
$u_{\eps} \in {\cal C}^{\infty}(\R)$ with properties
(\ref{properties-u}) and there is $C > 0$ such that
\begin{equation}
\label{bound-excited-state} \left\| u_{\eps} - \eta_{\eps}
\tanh\left(\frac{\cdot}{\sqrt{2} \eps}\right) \right\|_{L^{\infty}}
\leq C \eps^{2/3}.
\end{equation}
In particular, the solution converges pointwise
as $\eps \to 0$ to
$$
u_0(x) := \lim_{\eps \to 0} u_{\eps}(x) = \eta_0(x) {\rm sign}(x),
\quad x \in \mathbb{R}.
$$
\label{proposition-excited-state}
\end{theorem}

\begin{remark}
Function $v_{\eps}(x) = \tanh\left(\frac{x}{\sqrt{2} \eps}\right)$
is termed as the dark soliton. It is a solution of the
second-order equation
$$
\eps^2 v_{\eps}''(x) + (1 - v_{\eps}^2(x)) v_{\eps}(x) = 0, \quad
x \in \R,
$$
which arises in the context of the defocusing nonlinear
Schr\"{o}dinger equation.   \label{remark-dark-soliton}
\end{remark}

The proof of Theorem \ref{proposition-excited-state} consists of
six steps.

{\bf Step 1: Decomposition.} Let us substitute $u_{\eps}(x) =
\eta_{\eps}(x) \tanh\left(\frac{x}{\sqrt{2} \eps}\right) +
w_{\eps}(x)$ to the stationary equation (\ref{stationaryGPexc})
and obtain an equivalent problem for $w_{\eps}$ written in the
operator form
\begin{equation}
\label{operator-form} L_{\eps} w_{\eps} = H_{\eps} + N_{\eps}(w_{\eps}),
\end{equation}
where
$$
L_{\eps} := -\eps^2 \partial_x^2 + x^2 - 1 + 3 \eta_{\eps}^2(x) \tanh^2\left( \frac{x}{\sqrt{2} \eps} \right),
$$
$$
H_{\eps}(x) := \eta_{\eps}(x) \left( \eta_{\eps}^2(x) - 1 \right)
{\rm sech}^2\left( \frac{x}{\sqrt{2} \eps} \right)
\tanh\left( \frac{x}{\sqrt{2} \eps} \right)  + \sqrt{2} \eps
\eta_{\eps}'(x) {\rm sech}^2\left( \frac{x}{\sqrt{2} \eps}
\right),
$$
and
$$
N_{\eps}(w_{\eps})(x) = - 3 \eta_{\eps}(x)
\tanh\left( \frac{x}{\sqrt{2} \eps} \right) w_{\eps}^2(x) -
w_{\eps}^3(x).
$$
Let $x = \sqrt{2} \eps z$, where $z \in \R$ is a new variable, and
denote
$$
\hat{\eta}_{\eps}(z) := \eta_{\eps}(\sqrt{2} \eps z), \;\;
\hat{w}_{\eps}(z) := w_{\eps}(\sqrt{2} \eps z), \;\;
\hat{H}_{\eps}(z) := H_{\eps}(\sqrt{2} \eps z), \;\;
\hat{N}_{\eps}(\hat{w}_{\eps})(z) := N_{\eps}(w_{\eps})(\sqrt{2} \eps z).
$$

{\bf Step 2: Linear estimates.} In new variables,
operator $L_{\eps}$ becomes
$$
\hat{L}_{\eps} = -\frac{1}{2} \partial_z^2 + 2 \eps^2 z^2 - 1 +
3 \hat{\eta}_{\eps}^2(z) \tanh^2(z) = \hat{L}_0 + \hat{U}_{\eps}(z),
$$
where
$$
\hat{L}_0 := -\frac{1}{2} \partial_z^2 + 2 - 3 {\rm sech}^2(z)
$$
and
$$
\hat{U}_{\eps}(z) := 2 \eps^2 z^2 + 3 (\hat{\eta}_{\eps}^2(z) - 1) \tanh^2(z).
$$

Operator $\hat{L}_0$ is well known in the linearization of the
defocusing NLS equation at the dark soliton. The spectrum of
$\hat{L}_0$ in $L^2(\R)$ consists of two eigenvalues at $0$ and
$\frac{3}{2}$ with eigenfunctions ${\rm sech}^2(z)$ and $\tanh(z)
{\rm sech}(z)$ and the continuous spectrum on $[2,\infty)$. For
any $\hat{f} \in L^2_{\rm odd}(\R)$, there exists a unique
$\hat{L}_0^{-1} \hat{f} \in H^2_{\rm odd}(\R)$ such that
\begin{equation}
\label{bound-resolvent}
\exists C > 0 : \quad \forall \hat{f} \in L^2_{\rm odd}(\R) : \quad \| \hat{L}_0^{-1} \hat{f} \|_{H^2}
\leq C \| \hat{f} \|_{L^2}.
\end{equation}
Let us consider functions that decay to zero as $|z| \to \infty$ with
a fixed exponential decay rate $\alpha > 0$. Let $L^{\infty}_{\alpha}(\R)$
be the exponentially weighted space with the supremum norm
$$
\| \hat{w}_{\eps} \|_{L^{\infty}_{\alpha}} :=  \| e^{\alpha |\cdot|}
\hat{w}_{\eps} \|_{L^{\infty}}.
$$
The unique solution $\hat{L}_0^{-1} \hat{f}$ for any $\hat{f} \in
L^2_{\rm odd}(\R)$ is expressed explicitly by the integral formula
$$
\hat{L}_0^{-1} \hat{f}(z) = -2 {\rm sech}^2(z) \int_0^z
\cosh^4(z') \left( \int_{-\infty}^{z'} \hat{f}(z'') {\rm sech}^2(z'') dz''
\right) dz'.
$$
For any fixed $\alpha > 0$, it follows from the integral
representation that the solution $\hat{L}_0^{-1} \hat{f}$ decays
exponentially with the same rate as $\hat{f}$ so that
\begin{equation}
\label{bound-resolvent-weight} \exists C > 0 : \quad \forall
\hat{f} \in L^2_{\rm odd}(\R) \cap L^{\infty}_{\alpha}(\R) : \quad
\| \hat{L}_0^{-1} \hat{f} \|_{L^{\infty}_{\alpha}} \leq C \|
\hat{f} \|_{L^{\infty}_{\alpha}}.
\end{equation}

Figure 1 shows the confining potential $V_{\eps}(x) = x^2 - 1 + 3
\eta_{\eps}^2(x) \tanh^2(z)$ of operator $L_{\eps} = -\eps^2
\partial_x^2 + V_{\eps}(x)$ (solid line) and the bounded
potential $V_0(x) = -1 + 3 \tanh^2(z)$ of operator $L_0 = -\eps^2 \partial_x^2 + V_0(x)$ (dots)
versus $x$. The confining potential $V_{\eps}(x)$ has two wells near $x =
\pm 1$ and a deeper central well near $x = 0$. The two wells near $x = \pm 1$ are
absent in the potential $V_0(x)$.

Because of the confining potential, the spectrum of $\hat{L}_{\eps}$ is
purely discrete (Theorem 10.7 in \cite{Sigal}). It contains small eigenvalues 
that correspond to eigenfunctions localized in the central well near $z = 0$
and in the two smaller wells near $z = \pm \frac{1}{\sqrt{2} \eps}$.

We note that a similar operator at the ground state $\eps_{\eps}$
$$
\tilde{L}_{\eps} = -\eps^2 \partial_x^2 + x^2 - 1 + 3 \eta_{\eps}^2(x)
$$
was studied by Gallo \& Pelinovsky \cite{GalPel2}, where it was shown
that $\tilde{V}_{\eps}(x) = x^2 - 1 + 3 \eta_{\eps}^2(x) > 0$ for all $x \in \R$.
By property (P4), $\tilde{V}_{\eps}(x)$ is bounded away from zero
near $x = \pm 1$ by the constant of the order of ${\cal O}(\eps^{2/3})$. As a consequence, 
the purely discrete spectrum of $\tilde{L}_{\eps}$ in $L^2_{\rm odd}(\R)$ 
includes small positive eigenvalues of the order ${\cal O}(\eps^{2/3})$ with the eigenfunctions
localized in the two wells near $x = \pm 1$ (see Theorem 2 in \cite{GalPel2}). 

Thanks to the proximity of $\tanh^2(z)$ to $1$ near $z = \pm \frac{1}{\sqrt{2} \eps}$ with
an exponential accuracy in $\eps$, the potential $V_{\eps}(x)$ is similar to $\tilde{V}_{\eps}(x)$
near $x = \pm 1$ and satisfies for any fixed $x_0 > 0$:
$$
\exists C > 0 : \quad V(x) \geq C \eps^{2/3}, \quad |x| \geq x_0.
$$
On the other hand, for any fixed $z_0 > 0$, property (P2) implies that
$$
\exists C > 0 : \quad \sup_{|z| \leq z_0} |\hat{U}_{\eps}(z)| \leq
C \eps^2.
$$
Thanks to the positivity of $V_{\eps}(x)$ near $x = \pm 1$ and the proximity of
the central well near $x = 0$ in the potentials $V_{\eps}(x)$ and $V_0(x)$,
the quantum tunneling theory \cite{Sigal} implies that
the simple zero eigenvalue of $\hat{L}_0$ persists as a small eigenvalue of $\hat{L}_{\eps}$.
This eigenvalue of $\hat{L}_{\eps}$ corresponds to an even eigenfunction.
The other eigenvalue of $\hat{L}_0$ corresponding to an odd eigenfunction is bounded away from zero.

All other eigenvalues of $\hat{L}_{\eps}$ are small positive of the size ${\cal O}(\eps^{2/3})$.
As a result, operator $\hat{L}_{\eps}$ is still invertible
on $L^2_{\rm odd}(\R)$ but bound (\ref{bound-resolvent}) is now replaced by
\begin{equation}
\label{bound-resolvent-modified}
\exists C > 0 : \quad \forall \hat{f} \in L^2_{\rm odd}(\R) : \quad
\| \hat{L}_{\eps}^{-1} \hat{f} \|_{H^2} \leq C \eps^{-2/3} \| \hat{f} \|_{L^2}.
\end{equation}
Note that the function $\hat{L}_{\eps}^{-1} \hat{f} \in H^2_{\rm odd}(\R)$
has peaks near points $z = \pm \frac{1}{\sqrt{2} \eps}$ and $z = 0$.

\begin{figure}
\begin{center}
\includegraphics[width=0.4\textwidth]{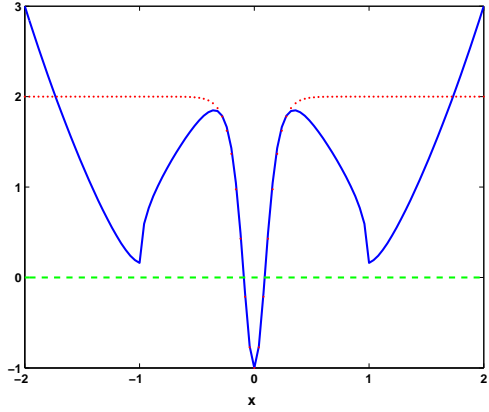}
\end{center}
\caption{Potentials of operators $L_{\eps}$ (solid line) and $L_0$ (dots) for the first excited state.}
\end{figure}

{\bf Step 3: Bounds on the inhomogeneous and nonlinear terms.} By symmetries, we note
that
$$
 \hat{H}_{\eps} \in L^2_{\rm odd}(\R) \quad \mbox{\rm and} \quad
\hat{N}_{\eps}(\hat{w}_{\eps}) : H^2_{\rm odd}(\R) \mapsto L^2_{\rm odd}(\R).
$$
We will show that for small $\eps > 0$ and fixed $\alpha \in (0,2)$ there is $C > 0$ such that
\begin{equation}
\label{first-bound-H}
\| \hat{H}_{\eps} \|_{L^2 \cap L^{\infty}_{\alpha}} \leq C \eps^{2/3}.
\end{equation}
Using the triangle inequality, we obtain
\begin{eqnarray*}
& \phantom{t} &  \| \hat{H}_{\eps} \|_{L^2} \leq \| \eta_{\eps} \|_{L^{\infty}}
\| (1 - \hat{\eta}^2_{\eps}) {\rm sech}^2(\cdot) \|_{L^2}
+ \sqrt{2} \eps \| \eta_{\eps}' \|_{L^{\infty}} \| {\rm sech}^2(\cdot) \|_{L^2}.
\end{eqnarray*}
By properties (P1) and (P2), for small $\eps > 0$ and fixed $\alpha \in (0,2)$
the first term is estimated by
\begin{eqnarray*}
\| (1 - \hat{\eta}^2_{\eps}) {\rm sech}^2(\cdot) \|_{L^2}  & \leq &
\| (1 - \hat{\eta}^2_{\eps}) {\rm sech}^2(\cdot) \|_{L^2(|z| \leq \eps^{-1/3})}
+ \| (1 - \hat{\eta}^2_{\eps}) {\rm sech}^2(\cdot) \|_{L^2(|z| \geq \eps^{-1/3})} \\
& \leq & \| 1 - \eta_{\eps}^2 \|_{L^{\infty}(|x| < \sqrt{2} \eps^{2/3})} \| {\rm sech}^2(\cdot) \|_{L^2}
+ \alpha^{-1/2} e^{-\alpha \eps^{-1/3}} \| {\rm sech}^2(\cdot) \|_{L^{\infty}_{\alpha}} \\
& \leq & C \eps^{4/3}.
\end{eqnarray*}
By property (P3), the second term is estimated by $C \eps^{2/3}$. As a result,
for any small $\eps > 0$ there is $C > 0$ such that $\|
\hat{H}_{\eps} \|_{L^2} \leq C \eps^{2/3}$. By similar arguments,
$\hat{H}_{\eps} \in L^{\infty}_{\alpha}(\R)$ for any $\alpha \in
(0,2)$ and there is $C > 0$ such that $\| \hat{H}_{\eps}
\|_{L^{\infty}_{\alpha}} \leq C \eps^{2/3}$.

To deal with the nonlinear terms, we recall that $H^2(\R)$ is Banach algebra with
respect to multiplication in the sense that
$$
\forall \hat{u},\hat{v} \in H^2(\R) : \quad \| \hat{u} \hat{v} \|_{H^2} \leq
\| \hat{u} \|_{H^2} \| \hat{v} \|_{H^2}
$$
For any $\hat{w}_{\eps} \in H^2(\R)$, we have
\begin{eqnarray}
\label{bound-on-N-1}
\| \hat{N}_{\eps}(\hat{w}_{\eps}) \|_{L^2} \leq 3 \| \eta_{\eps} \|_{L^{\infty}} \| \hat{w}_{\eps}^2 \|_{H^2} + \| \hat{w}_{\eps}^3 \|_{H^2} \leq 3 \| \hat{w}_{\eps} \|^2_{H^2}
+ \| \hat{w}_{\eps} \|^3_{H^2}.
\end{eqnarray}
Similarly, $L^{\infty}_{\alpha}(\R)$ is a Banach algebra with respect to multiplication for any
$\alpha \geq 0$.

{\bf Step 4: Normal-form transformations.} Because we are going to lose $\eps^{2/3}$
as a result of bound (\ref{bound-resolvent-modified}), we need to perform transformations
of solution $\hat{w}_{\eps}$, usually referred to as the normal-form transformations.
We need two normal-form transformations to ensure that the resulting operator
of a fixed-point equation is a contraction.

Let
$$
\hat{w}_{\eps} = \hat{w}_1 + \hat{w}_2 + \hat{\varphi}_{\eps}, \quad \hat{w}_1 = \hat{L}_0^{-1} \hat{H}_{\eps},
\quad \hat{w}_2 = -3 \hat{L}_0^{-1} \hat{\eta}_{\eps} \tanh(z) \hat{w}_1^2.
$$
The remainder term $\hat{\varphi}_{\eps}$ solves the new problem
\begin{equation}
\label{operator-form-new}
{\cal L}_{\eps} \hat{\varphi}_{\eps} = {\cal H}_{\eps} +
{\cal N}_{\eps}(\hat{\varphi}_{\eps}),
\end{equation}
where the new linear operator is
$$
{\cal L}_{\eps} := \hat{L}_{\eps} + \Delta \hat{U}_{\eps}(z), \quad
\Delta \hat{U}_{\eps}(z) := 6 \hat{\eta}_{\eps} \tanh(z)
(\hat{w}_1 + \hat{w}_2) + 3 (\hat{w}_1 + \hat{w}_2)^2,
$$
the new source term is
$$
{\cal H}_{\eps} := -\hat{U}_{\eps} (\hat{w}_1 + \hat{w}_2) - 3
\hat{\eta}_{\eps} \tanh(z) (2 \hat{w}_1 \hat{w}_2 + \hat{w}_2^2) -
(\hat{w}_1 + \hat{w}_2)^3,
$$
and the new nonlinear function is
$$
{\cal N}_{\eps}(\hat{\varphi}_{\eps}) := - 3 \hat{\eta}_{\eps}
\tanh(z) \hat{\varphi}^2_{\eps} - 3( \hat{w}_1 + \hat{w}_2)
\hat{\varphi}_{\eps}^2 - \hat{\varphi}_{\eps}^3.
$$

Thanks to bounds (\ref{bound-resolvent}),
(\ref{bound-resolvent-weight}), and (\ref{first-bound-H}),
we have $\hat{w}_1,\hat{w}_2 \in
H^2_{\rm odd}(\R) \cap L^{\infty}_{\alpha}(\R)$ for fixed $\alpha
\in (0,2)$ and
\begin{equation}
\label{normal-form-bound}
\exists C > 0 : \quad \| \hat{w}_1 \|_{H^2 \cap
L^{\infty}_{\alpha}} \leq C \eps^{2/3}, \quad \| \hat{w}_2 \|_{H^2 \cap
L^{\infty}_{\alpha}} \leq C \eps^{4/3}.
\end{equation}
As a result, for any small $\eps > 0$, there is $C > 0$ such that
$$
\| \hat{\eta}_{\eps} \tanh(z) (2 \hat{w}_1 \hat{w}_2 +
\hat{w}_2^2) \|_{L^2} \leq C \eps^2, \quad \| (\hat{w}_1 +
\hat{w}_2)^3 \|_{L^2} \leq C \eps^2.
$$
Let us now estimate the term $\hat{U}_{\eps} (\hat{w}_1 + \hat{w}_2)$ in $L^2(\R)$.
By properties (P1) and (P2), for small $\eps > 0$ and fixed $\alpha \in (0,2)$
there are constants $C(\alpha),\tilde{C}(\alpha) > 0$ such that
\begin{eqnarray*}
\| \hat{U}_{\eps} \hat{w}_j \|_{L^2}  & \leq & 2 \eps^2 \| z^2 \hat{w}_j \|_{L^2} +
3 \| (\hat{\eta}_{\eps}^2 - 1) \hat{w}_j \|_{L^2} \\
& \leq & \eps^2 C(\alpha) \| \hat{w}_j \|_{L^{\infty}_{\alpha}}
+ 3 \| (1 - \hat{\eta}^2_{\eps}) \hat{w}_j \|_{L^2(|z| \leq \eps^{-1/3})}
+ 3 \| (1 - \hat{\eta}^2_{\eps})\hat{w}_j \|_{L^2(|z| \geq \eps^{-1/3})} \\
& \leq & \eps^2 C(\alpha) \| \hat{w}_j \|_{L^{\infty}_{\alpha}}
+ 3 \| 1 - \eta_{\eps}^2 \|_{L^{\infty}(|x| < \sqrt{2} \eps^{2/3})} \| \hat{w}_j \|_{L^2} + 3 \alpha^{-1/2}
e^{-\alpha \eps^{-1/3}} \| \hat{w}_j \|_{L^{\infty}_{\alpha}} \\
& \leq & \tilde{C}(\alpha) \eps^{4/3} \| \hat{w}_j \|_{L^2 \cap L^{\infty}_{\alpha}}, \quad
j = 1,2.
\end{eqnarray*}
In view of bound (\ref{normal-form-bound}), for any small $\eps > 0$
there is $C > 0$ such that
\begin{equation}
\label{normal-form-bound-new}
\| \hat{U}_{\eps} (\hat{w}_1 + \hat{w}_2) \|_{L^2}  \leq C \eps^2.
\end{equation}
Combining all together, we have established that ${\cal H}_{\eps}
\in L^2_{\rm odd}(\R)$ and for any small $\eps > 0$, there is $C >
0$ such that
\begin{equation}
\label{H-bound-new} \| {\cal H}_{\eps} \|_{L^2} \leq C \eps^2.
\end{equation}

For the nonlinear term, we still have ${\cal
N}_{\eps}(\hat{\varphi}_{\eps}) : H^2_{\rm odd}(\R) \mapsto
L^2_{\rm odd}(\R)$. Thanks to bound (\ref{normal-form-bound}), for
any $\hat{\varphi}_{\eps} \in B_{\delta}(H^2_{\rm odd})$ in the
ball of radius $\delta > 0$, for any small $\eps > 0$ there is $C(\delta) > 0$ such that
\begin{equation}
\label{nonlinear-term-bound} \| {\cal
N}_{\eps}(\hat{\varphi}_{\eps}) \|_{L^2} \leq C(\delta) \|
\hat{\varphi}_{\eps} \|_{H^2}^2.
\end{equation}
Similarly, we obtain that ${\cal N}_{\eps}$ is Lipschitz continuous
in the ball $B_{\delta}(H^2_{\rm odd})$ and for any small $\eps > 0$ there is $C(\delta) > 0$
such that
\begin{equation}
\label{nonlinear-term-bound-lipschitz} \forall \hat{\varphi}_{\eps},
\hat{\phi}_{\eps} \in B_{\delta}(H^2_{\rm odd}) : \quad \| {\cal
N}_{\eps}(\hat{\varphi}_{\eps}) - {\cal
N}_{\eps}(\hat{\phi}_{\eps})\|_{L^2} \leq C(\delta) \left( \|
\hat{\varphi}_{\eps} \|_{H^2} + \| \hat{\phi}_{\eps} \|_{H^2} \right)
\| \hat{\varphi}_{\eps} - \hat{\phi} \|_{H^2}.
\end{equation}

{\bf Step 5: Fixed-point arguments.} Thanks to bound (\ref{normal-form-bound})
and Sobolev embedding of $H^2(\R)$ to $L^{\infty}(\R)$, $|\Delta \hat{U}_{\eps}(z)|$
is as small as ${\cal O}(\eps^{2/3})$ in the central well near $z = 0$ and is
exponentially small in $\eps$ in the two wells near $z = \pm  \frac{1}{\sqrt{2} \eps}$.
As a result, small positive eigenvalues of $\hat{L}_{\eps}$ of the size
${\cal O}(\eps^{2/3})$ persist in the spectrum of ${\cal L}_{\eps}$ and have the same
size, so that bound (\ref{bound-resolvent-modified}) extends to
operator ${\cal L}_{\eps}$ in the form
\begin{equation}
\label{bound-resolvent-final}
\exists C > 0 : \quad \forall \hat{f} \in L^2_{\rm odd}(\R) : \quad
\| {\cal L}_{\eps}^{-1} \hat{f} \|_{H^2} \leq C \eps^{-2/3} \| \hat{f} \|_{L^2}.
\end{equation}

Let us rewrite
equation (\ref{operator-form-new}) as the fixed-point problem
\begin{equation}
\label{fixed-point-problem}
\hat{\varphi}_{\eps} \in H^2_{\rm odd}(\R) : \quad
\hat{\varphi}_{\eps} = {\cal L}_{\eps}^{-1} {\cal H}_{\eps} +
{\cal L}_{\eps}^{-1} {\cal N}_{\eps}(\hat{\varphi}_{\eps}).
\end{equation}
The map $\hat{\varphi}_{\eps} \mapsto {\cal L}_{\eps}^{-1} {\cal
N}_{\eps}(\hat{\varphi}_{\eps})$ is Lipschitz continuous in the
neighborhood of $0 \in H^2_{\rm odd}(\R)$. Thanks to bounds
(\ref{nonlinear-term-bound}) and (\ref{bound-resolvent-final}),
the map is a contraction in the ball $B_{\delta}(H^2_{\rm odd})$
if $\delta \ll \eps^{2/3}$. On the other hand, thanks to bounds
(\ref{H-bound-new}) and (\ref{bound-resolvent-final}), the source term ${\cal L}_{\eps}^{-1} {\cal
H}_{\eps}$ is as small as ${\cal O}(\eps^{4/3})$ in $L^2$ norm. By
Banach's Fixed-Point Theorem in the ball $B_{\delta}(H^2_{\rm odd})$ with
$\delta \sim \eps^{4/3}$, there exists a unique 
$\hat{\varphi}_{\eps} \in H^2_{\rm odd}(\R)$ of
the fixed-point problem (\ref{fixed-point-problem}) such that
$$
\exists C > 0 : \quad \| \hat{\varphi}_{\eps} \|_{H^2} \leq C \eps^{4/3}.
$$
By Sobolev's embedding of $H^2(\R)$ to ${\cal C}^1(\R)$, for any small
$\eps > 0$ there is $C > 0$ such that
$$
\| w_{\eps} \|_{L^{\infty}} = \| \hat{w}_{\eps} \|_{L^{\infty}}
\leq C \| \hat{w}_1 + \hat{w}_2 + \hat{\varphi}_{\eps} \|_{H^2} \leq C \eps^{2/3},
$$
which completes the proof of bound (\ref{bound-excited-state}).

{\bf Step 6: Properties (\ref{properties-u}).} Solution
$\hat{w}_{\eps}$ constructed in Step (5) is a odd continuously differentiable
function of $z$ on $\R$ vanishing at infinity, so that
$u_{\eps}(0) = 0$ and $\lim_{x \to \infty} u_{\eps}(x) = 0$.
By bootstrapping arguments for the stationary equation (\ref{stationaryGPexc}),
we have $u_{\eps} \in {\cal C}^{\infty}(\R)$. It
remains to prove that $u_{\eps}(x)$ is positive for all $x \in
\R_+$.

Recall that $\eta_{\eps}(x) > 0$ for all $x \in \R$. By property
(P4) and bound (\ref{bound-excited-state}), there is $C > 0$ such
that $u_{\eps}(x) \geq C \eps^{1/3}$ for all $x \in [1,\sqrt{1 +
\eps^{2/3}}]$. We shall prove that $u_{\eps}(x) > 0$ for all $x
\geq \sqrt{1 + \eps^{2/3}}$. Assume by contradiction that there is
$x_0 > \sqrt{1 + \eps^{2/3}}$ such that $u_{\eps}(x_0) = 0$ and
$u_{\eps}'(x_0) < 0$. (If $u_{\eps}'(x_0) = 0$, then $u_{\eps}(x) = 0$ is the only solution
of the second-order equation (\ref{stationaryGPexc}).)
The continuity of $u_{\eps}(x)$
implies that $u_{\eps}(x) < 0$ for every $x \in (x_0,\tilde{x}_0)$
for some $\tilde{x}_0 > x_0$. Using the differential equation
(\ref{stationaryGPexc}), we obtain
$$
u_{\eps}''(x) = \frac{1}{\eps^2} (x^2 - 1 + u^2_{\eps}(x)) u_{\eps}(x) < 0, \quad x \in (x_0,\tilde{x}_0).
$$
Then, $u_{\eps}'(x) \leq u_{\eps}'(x_0) < 0$, so that
$u_{\eps}(x)$ is a negative, decreasing function of $x$ for all $x
> x_0$ with $\tilde{x}_0 = \infty$. This fact is a contradiction
with the decay of $u_{\eps}(x)$ to zero as $x \to \infty$.
Therefore, $u_{\eps}(x) > 0$ for all $x \in \R_+$.

Combining results in Steps (5) and (6), we conclude that
$u_{\eps}(x)$ is the first excited state of the stationary
equation (\ref{stationaryGPexc}) that satisfies properties
(\ref{properties-u}).

\section{Second excited state}

The second excited state is an even solution of the stationary
equation (\ref{stationaryGPexc}) such that
\begin{equation}
\label{properties-uu} u_{\eps}(x) > 0 \;\; \mbox{\rm for all} \;\; |x| > x_0, \quad
u_{\eps}(x) < 0 \;\; \mbox{\rm for all} \;\; |x| < x_0, \quad \mbox{\rm
and} \quad \lim_{x \to \infty} u_{\eps}(x) = 0.
\end{equation}
Here $x_0 > 0$ determines a location of two symmetric zeros of
$u_{\eps}(x)$ at $x = \pm x_0$. The second excited state is
approximated as $\eps \to 0$ by a product of two copies of dark
solitons (Remark \ref{remark-dark-soliton}) placed at $x = \pm a$
with $a \approx x_0$ as $\eps \to 0$. Our analysis is based on the
method of Lyapunov--Schmidt reductions, which gives existence and
convergence properties for the second excited state, as well as an
analytical expansion of $a$ for small $\eps$.

\begin{theorem}
For sufficiently small $\eps > 0$, there exists a unique solution
$u_{\eps} \in {\cal C}^{\infty}(\R)$ with properties (\ref{properties-uu})
and there exist $a > 0$ and $C > 0$ such that
\begin{equation}
\label{bound-second-excited-state} \left\| u_{\eps} - \eta_{\eps}
\tanh\left(\frac{\cdot - a}{\sqrt{2} \eps}\right)
\tanh\left(\frac{\cdot + a}{\sqrt{2} \eps}\right) \right\|_{L^{\infty}}
\leq C \eps^{2/3}
\end{equation}
and
\begin{equation}
\label{bound-a-asymptotic} a = - \frac{\eps}{\sqrt{2}} \left(
\log(\eps) + \frac{1}{2} \log|\log(\eps)|
 - \frac{3}{2} \log(2) + o(1) \right) \quad \mbox{\rm as} \quad \eps \to 0.
\end{equation}
\label{proposition-second-excited-state}
\end{theorem}

\begin{remark}
Since $a \to 0$ as $\eps \to 0$ while $\eta_{\eps}(x) \approx 1$
near $x = 0$, we have
$$
x_0 = a + {\cal O}(\eps^{5/3}) \quad \mbox{\rm as} \quad \eps \to
0.
$$ \label{remark-zeros}
\end{remark}

\begin{remark}
Exactly the same asymptotic expansion (\ref{bound-a-asymptotic})
has been obtained with the use of the averaged Lagrangian
approximation and has been confirmed numerically \cite{CKP}.
\end{remark}

The proof of Theorem \ref{proposition-second-excited-state}
follows the same steps as the proof of Theorem
\ref{proposition-excited-state} with an additional step on the
Lyapunov--Schmidt bifurcation equation.

{\bf Step 1: Decomposition.} Let $a \in (0,1)$ and substitute
$$
u_{\eps}(x) = \eta_{\eps}(x) \tanh\left(\frac{x - a}{\sqrt{2} \eps}\right)
\tanh\left(\frac{x + a}{\sqrt{2} \eps}\right) + w_{\eps}(x)
$$
to the stationary equation (\ref{stationaryGPexc}). The equivalent problem
for $w_{\eps}$ takes the operator form
\begin{equation}
\label{operator-form-second} L_{\eps} w_{\eps} = H_{\eps} + N_{\eps}(w_{\eps}),
\end{equation}
where
$$
L_{\eps} := -\eps^2 \partial_x^2 + x^2 - 1 + 3 \eta_{\eps}^2(x)
\tanh^2(z_+) \tanh^2(z_-),
$$
\begin{eqnarray*}
H_{\eps} & := & \eta_{\eps}(x) (\eta_{\eps}^2(x) - 1) \tanh(z_+) \tanh(z_-) \left(
{\rm sech}^2(z_+) + {\rm sech}^2(z_-) \right) \\
& \phantom{t} & + \eta_{\eps}(x) {\rm sech}^2(z_+) {\rm sech}^2(z_-) \left( 1 - \eta_{\eps}^2(x)
\tanh(z_+) \tanh(z_-) \right) \\
& \phantom{t} & + \sqrt{2} \eps \eta_{\eps}'(x) \left( \tanh(z_+)
{\rm sech}^2(z_-) +  \tanh(z_-) {\rm sech}^2(z_+) \right),
\end{eqnarray*}
and
\begin{eqnarray*}
N_{\eps}(w_{\eps}) = - 3 \eta_{\eps}(x) \tanh(z_+) \tanh(z_-)
w_{\eps}^2(x) - w_{\eps}^3(x),
\end{eqnarray*}
with the following notations
$$
z_{\pm} = z \pm \zeta, \quad z = \frac{x}{\sqrt{2} \eps}, \quad
\zeta = \frac{a}{\sqrt{2} \eps}.
$$
We again denote the functions in $z$ by hats. We shall assume a
priori that
\begin{equation}
\label{bounds-on-a} \left\{ \begin{array}{l} \exists \beta \in (0,1) : \quad a \leq \sqrt{2} \beta \eps^{2/3}, \\
\exists C > 0 : \quad e^{-2 \zeta} \leq C \eps |\log(\eps)|^{1/2}.
\end{array} \right.
\end{equation}
Note that bounds (\ref{bounds-on-a}) imply that
$a \to 0$ and $\zeta \to \infty$ as $\eps \to 0$.

{\bf Step 2: Linear estimates.} In new variables, operator
$L_{\eps}$ becomes
$$
\hat{L}_{\eps} = -\frac{1}{2} \partial_z^2 + 2 \eps^2 z^2 - 1 + 3
\hat{\eta}_{\eps}^2(z) \tanh^2(z + \zeta) \tanh^2(z-\zeta) \equiv
\hat{L}_0(\zeta) + \hat{U}_{\eps}(z,\zeta),
$$
where
$$
\hat{L}_0(\zeta) = -\frac{1}{2} \partial_z^2 + 2 - 3 {\rm
sech}^2(z+\zeta) - 3 {\rm sech}^2(z - \zeta)
$$
and
$$
\hat{U}_{\eps}(z,\zeta) = 2 \eps^2 z^2 + 3 (\hat{\eta}^2_{\eps}(z)
- 1) \tanh^2(z + \zeta) \tanh^2(z-\zeta) + 3 {\rm sech}^2(z +
\zeta) {\rm sech}^2(z - \zeta).
$$

Operator $\hat{L}_0(\zeta)$ has now two eigenvalues in the
neighborhood of $0$ for large $\zeta$ because of the double-well
potential centered at $z = \pm \zeta$. If $\zeta$ is large, the geometric
splitting theory \cite{Sandstede} implies that the
eigenfunctions $\hat{\psi}^{\pm}_0(z)$ of operator
$\hat{L}_0(\zeta)$ corresponding to the two smallest eigenvalues are given
asymptotically by
\begin{equation}
\label{eigenfunction-1} \hat{\psi}^{\pm}_0(z) =
\frac{\hat{\psi}_0(z-\zeta) \pm \hat{\psi}_0(z+\zeta)}{\sqrt{2}}
+ {\cal O}_{L^{\infty}}(e^{-2 \zeta}) \quad
\mbox{\rm as} \quad \zeta \to \infty,
\end{equation}
where $\hat{\psi}_0(z) = \frac{\sqrt{3}}{2} {\rm sech}^2(z)$ is
the $L^2$-normalized eigenfunction of $\hat{L}_0 = -\frac{1}{2}
\partial_z^2 + 2 - 3 {\rm sech}^2(z)$ for the zero eigenvalue.

Note that $\hat{\psi}^+_0(z)$ is even and $\hat{\psi}^-_0(z)$ is odd
on $\R$. For the second excited state, we are looking for an even
solution $\hat{w}_{\eps}(z)$. Since $a$ is not specified yet, we
add the condition $\langle \hat{\psi}^+_0, \hat{w}_{\eps} \rangle =
0$ and define a constrained subspace of $H^2_{\rm even}(\R)$ by
$$
X_0 = \{ \hat{w}_{\eps} \in H^2_{\rm even}(\R) : \quad \langle
\hat{\psi}^+_0, \hat{w}_{\eps} \rangle = 0 \}.
$$
Let $P_0$ be an orthogonal projection operator to the complement
of $\hat{\psi}^+_0$ in $L^2_{\rm even}(\R)$. Since eigenfunction
$\hat{\psi}^-_0$ is odd and the rest of spectrum of
$\hat{L}_0(\zeta)$ is bounded from zero, for any $\hat{f} \in
L^2_{\rm even}(\R)$, there exists a unique $P_0
\hat{L}^{-1}_0(\zeta) P_0 \hat{f} \in H^2_{\rm even}(\R)$ such that
\begin{equation}
\label{resolvent-estimate-1} \exists C > 0 : \quad \forall \hat{f}
\in L^2_{\rm even}(\R) : \quad \| P_0 \hat{L}^{-1}_0(\zeta) P_0
\hat{f} \|_{H^2} \leq C \| \hat{f} \|_{L^2}.
\end{equation}

Let us consider functions that decay to zero as $|z - \zeta|, |z +
\zeta| \to \infty$ with a fixed exponential decay rate $\alpha >
0$. Let $L^{\infty}_{\alpha,\zeta}(\R)$ be the exponentially
weighted space with the supremum norm
$$
\| \hat{w}_{\eps} \|_{L^{\infty}_{\alpha,\zeta}} :=  \sup_{z \in \R_+}
e^{\alpha (|z - \zeta|)} |\hat{w}_{\eps}(z)| +
\sup_{z \in \R_-} e^{\alpha (|z + \zeta|)} |\hat{w}_{\eps}(z)|.
$$
For fixed $\alpha > 0$ and $\zeta > 0$, the unique solution $P_0
\hat{L}^{-1}_0(\zeta) P_0 \hat{f}$ decays exponentially with the
same rate as $\hat{f}$ so that
\begin{equation}
\label{resolvent-estimate-2} \exists C > 0 : \quad \forall \hat{f}
\in L^2_{\rm even}(\R) \cap L^{\infty}_{\alpha}(\R) : \quad \| P_0
\hat{L}^{-1}_0(\zeta) P_0 \hat{f}  \|_{L^{\infty}_{\alpha,\zeta}}
\leq C \| \hat{f} \|_{L^{\infty}_{\alpha,\zeta}}.
\end{equation}

Figure 2 shows the potential $V_{\eps}(x) = x^2 - 1 + 3
\eta_{\eps}^2(x) \tanh^2(z+\zeta) \tanh^2(z - \zeta)$ of operator
$L_{\eps} = -\eps^2 \partial_x^2 + V_{\eps}(x)$ (solid line) and
the potential $V_0(x) = 2 - 3 {\rm sech}^2(z+\zeta) - 3 {\rm
sech}^2(z - \zeta)$ of operator $L_0 = -\eps^2 \partial_x^2 +
V_0(x)$ (dots) versus $x$. The bounded potential $V_0(x)$ has two
wells near $x = \pm a$, whereas the confining
potential $V_{\eps}(x)$ has four wells
near $x = \pm a$ and $x = \pm 1$.

Again, the spectrum of operator $\hat{L}_{\eps}$ with a confining
potential is purely discrete. The two wells of the confining potential
$V_{\eps}(x)$ near $x = \pm 1$ are ${\cal O}(\eps^{2/3})$-close
to zero but still positive thanks to property (P4) and the fact
that $\tanh(z \pm \zeta) = 1$ with exponential accuracy in $\eps$ if
$\zeta \leq \beta \eps^{-1/3}$ for fixed $\beta \in (0,1)$. Therefore, for any fixed $x_0 > 0$,
we have
\begin{equation}
\label{estimate-1-potential}
\exists C > 0 : \quad V(x) \geq C \eps^{2/3}, \quad |x| \geq x_0.
\end{equation}
On the other hand, by property (P2) for any $\zeta \in (0,\eps^{-2/3})$,
we have
\begin{equation}
\label{estimate-2-potential}
\exists C > 0 : \quad \sup_{|z| \leq \eps^{-2/3}} |\hat{U}_{\eps}(z,\zeta)| \leq
C( \eps^{2/3} + e^{-4 \zeta} ).
\end{equation}
Thanks to properties (\ref{estimate-1-potential}) and (\ref{estimate-2-potential}),
the quantum tunneling theory \cite{Sigal} implies that
the two small eigenvalues of $\hat{L}_0$ persist as two small eigenvalues
of $\hat{L}_{\eps}$ with two eigenfunctions $\hat{\psi}^{\pm}_{\eps}$ that satisfy
asymptotically
\begin{equation}
\label{proximity-1}
\hat{\psi}^{\pm}_{\eps}(z) =
\hat{\psi}^{\pm}_0(z) + {\cal O}_{L^{\infty}}(\eps^{2/3}) \quad
\mbox{\rm as} \quad \eps \to 0,
\end{equation}
thanks to a priori bound (\ref{bounds-on-a}) and the exponential smallness of
$\hat{\psi}^{\pm}_0(z)$ in $\eps$ near $z = \pm \frac{1}{\sqrt{2} \eps}$.

Let $P_{\eps}$ be an orthogonal projection operator to the complement
of $\hat{\psi}^+_{\eps}$ in $L^2_{\rm even}(\R)$. Because of the small
${\cal O}(\eps^{2/3})$ eigenvalues of $\hat{L}_{\eps}$,
bound (\ref{resolvent-estimate-1}) is now replaced by
\begin{equation}
\label{resolvent-estimate-3}
\exists C > 0 : \quad \forall \hat{f} \in L^2_{\rm even}(\R) : \quad
\| P_{\eps} \hat{L}_{\eps}^{-1} P_{\eps} \hat{f} \|_{H^2} \leq C \eps^{-2/3} \| \hat{f} \|_{L^2}.
\end{equation}
The function $P_{\eps} \hat{L}_{\eps}^{-1} P_{\eps} \hat{f} \in H^2_{\rm even}(\R)$
has peaks in all four wells near points $z = \pm \frac{1}{\sqrt{2} \eps}$ and
$z = \pm \zeta$.

\begin{figure}
\begin{center}
\includegraphics[width=0.45\textwidth]{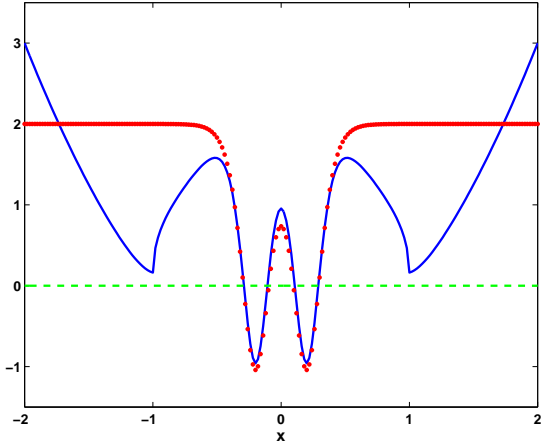}
\end{center}
\caption{Potential of operator $L_{\eps}$ (solid line) and $L_0$
(dots) for the second excited state.}
\end{figure}

{\bf Step 3: Bounds on the inhomogeneous and nonlinear terms.} From the symmetry of
terms in $\hat{H}_{\eps}$ and $\hat{N}_{\eps}(\hat{w}_{\eps})$, we have
$$
\hat{N}_{\eps}(\hat{w}_{\eps}) : H^2_{\rm even}(\R) \mapsto L^2_{\rm even}(\R) \quad
{\rm and} \quad \hat{H}_{\eps} \in L^2_{\rm even}(\R).
$$
Under a priori bound (\ref{bounds-on-a}), we first show that there
is $C > 0$ such that
\begin{eqnarray}
\label{bound-on-H} \| \hat{H}_{\eps} \|_{L^2} & \leq & C
\eps^{2/3}.
\end{eqnarray}

The upper bound for the first term in $\hat{H}_{\eps}$ involves estimates of
$$
I_1(z) := ( 1 - \hat{\eta}^2_{\eps}(z)) ({\rm sech}^2(z + \zeta) + {\rm sech}^2(z - \zeta)),
$$
which may create a problem since $\zeta \to \infty$ as $\eps \to 0$
and $\hat{\eta}_{\eps}(z) \to 0$ as $|z| \to \infty$.
By properties (P1) and (P2), for any $\alpha \in (0,2)$,
$\zeta \leq \beta \eps^{-1/3}$ for any $\beta \in (0,1)$,
and any small $\eps > 0$, there is constant $C > 0$ such that
\begin{eqnarray*}
\| I_1 \|_{L^2} & \leq & \| I_1 \|_{L^2(|z| \leq \eps^{-1/3})} + \| I_1 \|_{L^2(|z| \geq \eps^{-1/3})} \\
& \leq & \| 1 - \eta_{\eps}^2 \|_{L^{\infty}(|x| < \sqrt{2} \eps^{2/3})}
\| {\rm sech}^2(z_+) + {\rm sech}^2(z_-) \|_{L^2}  \\
& \phantom{t} & \phantom{texttext} + \alpha^{-1/2} e^{-\alpha (\eps^{-1/3}-\zeta)} \| {\rm sech}^2(z_+) + {\rm sech}^2(z_-) \|_{L^{\infty}_{\alpha,\zeta}} \\
& \leq & C \eps^{4/3}.
\end{eqnarray*}
Thus, the condition $\zeta \leq \beta \eps^{-1/3}$ from a priori bound (\ref{bounds-on-a})
is sufficient to keep $I_1$ small in $L^2$.

The upper bound for the second term in
$\hat{H}_{\eps}$ involves the estimate of the overlapping term
$$
I_2(z) :=  {\rm sech}^2(z_+) {\rm sech}^2(z_-).
$$
Under a priori bound (\ref{bounds-on-a}), this term is estimated by
\begin{eqnarray*}
\| I_2 \|_{L^2} & \leq & \left( \int_{\mathbb{R}} {\rm sech}^4(z + \zeta) {\rm sech}^4(z -
\zeta) dz \right)^{1/2} \\
& = & \left( 2 \int_{-\zeta}^{\infty} {\rm sech}^4(u) {\rm
sech}^4(u + 2 \zeta) du \right)^{1/2} \leq C e^{-2\zeta} \leq C
\eps |\log(\eps)|^{1/2}.
\end{eqnarray*}
The last term in $\hat{H}_{\eps}$ is proportional to $\eps \eta_{\eps}'$
and is handled with property (P3) to give (\ref{bound-on-H}). By similar arguments,
$\hat{H}_{\eps} \in L^{\infty}_{\alpha,\zeta}(\R)$ for any $\alpha \in
(0,2)$ and $\zeta \leq \beta \eps^{-1/3}$ for any $\beta \in (0,1)$ and
for any small $\eps > 0$ there is $C > 0$ such that $\| \hat{H}_{\eps}
\|_{L^{\infty}_{\alpha,\zeta}} \leq C \eps^{2/3}$.

The nonlinear terms in $\hat{N}_{\eps}(\hat{w}_{\eps})$
are handled with the Banach algebra of $H^2(\R)$, so we obtain
\begin{eqnarray}
\label{bound-on-N} \| \hat{N}_{\eps}(\hat{w}_{\eps}) \|_{L^2} \leq
3 \| \hat{w}_{\eps} \|^2_{H^2} + \| \hat{w}_{\eps}
\|^3_{H^2}.
\end{eqnarray}

{\bf Step 4: Normal-form transformations.} Unlike step (4)
in the proof of Theorem \ref{proposition-excited-state},
we need to perform a sequence of two normal-form transformations because the
orthogonal projection
operator to the one-dimensional subspace spanned by an even eigenfunction for the smallest
eigenvalue of $\hat{L}_0$ has to be changed to the projection operator associated
with an eigenfunction of a new linearization operator. For the sake of short notations, we
combine both normal-form transformations and write them together.

Let $\hat{w}_{\eps} = \hat{w}_1 + \hat{w}_2 + \hat{\varphi}_{\eps}$
with
\begin{eqnarray*}
\hat{w}_1 = P_0 \hat{L}_0^{-1} P_0 \hat{H}_{\eps}, \quad
\hat{w}_2 = P_0 \hat{L}_0^{-1} P_0 \hat{G}_{\eps},
\end{eqnarray*}
where
$$
\hat{G}_{\eps} := -3 \hat{\eta}_{\eps} \tanh(z + \zeta) \tanh(z - \zeta) \hat{w}_1^2 +
({\cal P}_{\eps} - P_0) \hat{H}_{\eps}
$$
and ${\cal P}_{\eps}$ is a new orthogonal projection operator introduced below.

The remainder term $\hat{\varphi}_{\eps}$ solves the new problem
\begin{equation}
\label{operator-form-new-2}
{\cal L}_{\eps} \hat{\varphi}_{\eps} = {\cal H}_{\eps}
+ {\cal N}_{\eps}(\hat{\varphi}_{\eps}) + {\cal S}_{\eps},
\end{equation}
where the new linear operator is
$$
{\cal L}_{\eps} := \hat{L}_{\eps} + \Delta \hat{U}_{\eps}(z), \quad
\Delta \hat{U}_{\eps}(z) := 6 \hat{\eta}_{\eps} \tanh(z+\zeta)
\tanh(z-\zeta) (\hat{w}_1 + \hat{w}_2) + 3 (\hat{w}_1 + \hat{w}_2)^2,
$$
the new source term is
\begin{eqnarray*}
{\cal H}_{\eps} := -\hat{U}_{\eps} (\hat{w}_1 + \hat{w}_2) - 3
\hat{\eta}_{\eps} \tanh(z+\zeta) \tanh(z-\zeta) (2 \hat{w}_1 \hat{w}_2 + \hat{w}_2^2) -
(\hat{w}_1 + \hat{w}_2)^3 + ({\cal P}_{\eps} - P_0) \hat{G}_{\eps},
\end{eqnarray*}
the new nonlinear function is
$$
{\cal N}_{\eps}(\hat{\varphi}_{\eps}) := - 3 \hat{\eta}_{\eps}
\tanh(z+\zeta) \tanh(z - \zeta) \hat{\varphi}^2_{\eps} - 3( \hat{w}_1 + \hat{w}_2)
\hat{\varphi}_{\eps}^2 - \hat{\varphi}_{\eps}^3,
$$
and the new one-dimensional projection is
$$
{\cal S}_{\eps} := (I - {\cal P}_{\eps}) (\hat{H}_{\eps} + \hat{G}_{\eps}).
$$

If $\hat{w}_1, \hat{w}_2 \in H^2(\R) \cap L^{\infty}_{\alpha,\zeta}(\R)$ satisfy
bounds (\ref{normal-form-bound-2}) below, then $\Delta \hat{U}_{\eps}(z)$ is
as small as ${\cal O}(\eps^{2/3})$ in the two wells near $z = \pm \zeta$
and is exponentially small in $\eps$ in
the two wells near $z = \pm \frac{1}{\sqrt{2} \eps}$. Let $\tilde{\psi}^{\pm}_{\eps}$ be the
eigenfunctions of ${\cal L}_{\eps}$ for the two eigenvalues continued from
the two smallest eigenvalues of $\hat{L}_0$. The proximity
of the potential wells and expansion (\ref{proximity-1}) imply that
\begin{equation}
\label{proximity-2}
\tilde{\psi}^{\pm}_{\eps}(z) =
\hat{\psi}^{\pm}_{\eps}(z) + {\cal O}_{L^{\infty}}(\eps^{2/3})
= \hat{\psi}^{\pm}_0(z) + {\cal O}_{L^{\infty}}(\eps^{2/3}) \quad
\mbox{\rm as} \quad \eps \to 0.
\end{equation}
Let ${\cal P}_{\eps}$ be an orthogonal projection operator to the complement
of $\tilde{\psi}^+_{\eps}$ in $L^2_{\rm even}(\R)$. Thanks to expansion (\ref{proximity-2}), we have
\begin{equation}
\label{proximity}
\exists C > 0 : \quad \| {\cal P}_{\eps} - P_0 \|_{L^2 \to L^2} \leq C \eps^{2/3}.
\end{equation}

Thanks to bounds (\ref{resolvent-estimate-1}),
(\ref{resolvent-estimate-2}), (\ref{bound-on-H}),
and (\ref{proximity}), we have $\hat{w}_1,\hat{w}_2 \in
H^2_{\rm even}(\R) \cap L^{\infty}_{\alpha,\zeta}(\R)$ for any $\alpha
\in (0,2)$ and $\zeta \leq \beta \eps^{-1/3}$ for any $\beta \in (0,1)$ such that
\begin{equation}
\label{normal-form-bound-2}
\exists C > 0 : \quad \| \hat{w}_1 \|_{H^2 \cap
L^{\infty}_{\alpha,\zeta}} \leq C \eps^{2/3}, \quad \| \hat{w}_2 \|_{H^2 \cap
L^{\infty}_{\alpha,\zeta}} \leq C \eps^{4/3}.
\end{equation}
As a result, for any small $\eps > 0$, there is $C > 0$ such that
\begin{eqnarray*}
\| \hat{\eta}_{\eps} \tanh(z+\zeta) \tanh(z - \zeta) (2 \hat{w}_1 \hat{w}_2 +
\hat{w}_2^2) \|_{L^2} & \leq & C \eps^2, \\ \| (\hat{w}_1 +
\hat{w}_2)^3 \|_{L^2} & \leq & C \eps^2, \\
\| ({\cal P}_{\eps} - P_0) \hat{G}_{\eps} \|_{L^2} & \leq & C \eps^2.
\end{eqnarray*}
Let us now estimate the term $\hat{U}_{\eps} \hat{w}_j$ in $L^2(\R)$ for
any $j = \{1,2\}$. By properties (P1) and (P2), for any $\alpha \in (0,2)$ and
$\zeta \leq \beta \eps^{-1/3}$ for any $\beta \in (0,1)$, and for any small $\eps > 0$, we have
\begin{eqnarray*}
\| \eps^2 z^2 \hat{w}_j \|_{L^2}  & \leq & C \eps^2 \zeta^2 \| \hat{w}_j
\|_{L^{\infty}_{\alpha,\zeta}} \leq C \eps^{4/3} \| \hat{w}_j \|_{L^{\infty}_{\alpha,\zeta}}, \\
\| (1 - \hat{\eta}_{\eps}^2) \hat{w}_j \|_{L^2} & \leq & \| (1 - \hat{\eta}^2_{\eps}) \hat{w}_j \|_{L^2(|z| \leq \eps^{-1/3})} + \| (1 - \hat{\eta}^2_{\eps})\hat{w}_j \|_{L^2(|z| \geq \eps^{-1/3})}  \\
& \leq & \| 1 - \eta_{\eps}^2 \|_{L^{\infty}(|x| < \sqrt{2} \eps^{2/3})} \| \hat{w}_j \|_{L^2} + \alpha^{-1/2}
e^{-\alpha (\eps^{-1/3}-\zeta)} \| \hat{w}_j \|_{L^{\infty}_{\alpha,\zeta}} \\
& \leq & C \eps^{4/3} \| \hat{w}_j \|_{L^2 \cap L^{\infty}_{\alpha,\zeta}},  \\
\| {\rm sech}^2(z_+) {\rm sech}^2(z_-) \hat{w}_{\eps} \|_{L^2}
& \leq & C e^{-4\zeta} \| \hat{w}_{\eps} \|_{L^2} \leq C \eps^2
|\log(\eps)| \| \hat{w}_{\eps} \|_{L^2}.
\end{eqnarray*}
In view of bound (\ref{normal-form-bound-2}), for any small $\eps > 0$,
there is $C > 0$ such that
\begin{equation}
\label{normal-form-bound-new-2}
\| \hat{U}_{\eps} (\hat{w}_1 + \hat{w}_2) \|_{L^2}  \leq C \eps^2.
\end{equation}
Combining all together, we have established that ${\cal H}_{\eps}
\in L^2_{\rm even}(\R)$ and for any small $\eps > 0$, there is $C >
0$ such that
\begin{equation}
\label{H-bound-new-2} \| {\cal H}_{\eps} \|_{L^2} \leq C \eps^2.
\end{equation}

For the nonlinear term, we still have ${\cal
N}_{\eps}(\hat{\varphi}_{\eps}) : H^2_{\rm even}(\R) \mapsto
L^2_{\rm even}(\R)$. Thanks to bound (\ref{normal-form-bound-2}), for
any $\hat{\varphi}_{\eps} \in B_{\delta}(H^2_{\rm even})$ in the
ball of radius $\delta > 0$ and for any small $\eps > 0$,
there is $C(\delta) > 0$ such that
\begin{equation}
\label{nonlinear-term-bound-2} \| {\cal
N}_{\eps}(\hat{\varphi}_{\eps}) \|_{L^2} \leq C(\delta) \|
\hat{\varphi}_{\eps} \|_{H^2}^2.
\end{equation}

{\bf Step 5: Fixed-point arguments.} Because $\Delta \hat{U}_{\eps}(z)$
is exponentially small in $\eps$ near $z = \pm \frac{1}{\sqrt{2} \eps}$,
small positive eigenvalues of $\hat{L}_{\eps}$ of the size ${\cal O}(\eps^{2/3})$
persist in the spectrum of ${\cal L}_{\eps}$ and have the same size.
As a result, bound (\ref{resolvent-estimate-3}) extends
to the operator ${\cal L}_{\eps}$ in the form
\begin{equation}
\label{resolvent-estimate-4}
\exists C > 0 : \quad \forall \hat{f} \in L^2_{\rm even}(\R) : \quad
\| {\cal P}_{\eps} \hat{L}_{\eps}^{-1} {\cal P}_{\eps} \hat{f} \|_{H^2}
\leq C \eps^{-2/3} \| \hat{f} \|_{L^2},
\end{equation}
where the new projection operator ${\cal P}_{\eps}$ is used. As a result,
we rewrite equation (\ref{operator-form-new-2}) as the fixed-point problem
\begin{equation}
\label{fixed-point-problem-2}
\hat{\varphi}_{\eps} \in H^2_{\rm even}(\R) : \quad
\hat{\varphi}_{\eps} = {\cal P}_{\eps} {\cal L}_{\eps}^{-1} {\cal P}_{\eps} \left( {\cal H}_{\eps} +
{\cal N}_{\eps}(\hat{\varphi}_{\eps}) + {\cal S}_{\eps} \right).
\end{equation}
subject to the Lyapunov--Schmidt bifurcation equation
\begin{equation}
\label{LSreductions}
{\cal F}_{\eps} := \langle \tilde{\psi}^+_{\eps},({\cal H}_{\eps} +
{\cal N}_{\eps}(\hat{\varphi}_{\eps}) + {\cal S}_{\eps}) \rangle_{L^2} =
\langle \tilde{\psi}^+_{\eps},(\hat{H}_{\eps} + \hat{G}_{\eps} + {\cal H}_{\eps} +
{\cal N}_{\eps}(\hat{\varphi}_{\eps})) \rangle_{L^2} = 0.
\end{equation}

The map $\hat{\varphi}_{\eps} \mapsto {\cal P}_{\eps} {\cal L}_{\eps}^{-1} {\cal P}_{\eps} {\cal
N}_{\eps}(\hat{\varphi}_{\eps})$ is Lipschitz continuous in the
neighborhood of $0 \in H^2_{\rm even}(\R)$. Thanks to bounds
(\ref{nonlinear-term-bound-2}) and (\ref{resolvent-estimate-4}),
the map is a contraction in the ball $B_{\delta}(H^2_{\rm even})$
if $\delta \ll \eps^{2/3}$. On the other hand, thanks to bounds
(\ref{H-bound-new-2}) and (\ref{resolvent-estimate-4}),
the source term ${\cal P}_{\eps} {\cal L}_{\eps}^{-1} {\cal P}_{\eps} {\cal
H}_{\eps}$ is as small as ${\cal O}(\eps^{4/3})$ in $L^2$ norm.
Furthermore, ${\cal P}_{\eps} {\cal S}_{\eps} = 0$.

By Banach's Fixed-Point Theorem in the ball $B_{\delta}(H^2_{\rm even})$ with
$\delta \sim \eps^{4/3}$, for any $(a,\zeta)$ satisfying a priori bounds
(\ref{bounds-on-a}) and sufficiently small $\eps > 0$,
there exists a unique $\hat{\varphi}_{\eps} \in H^2_{\rm even}(\R)$ of
the fixed-point problem (\ref{fixed-point-problem-2}) and $C > 0$ such that
$$
\| \hat{\varphi}_{\eps} \|_{H^2} \leq C \eps^{4/3}.
$$
By Sobolev's embedding of $H^2(\R)$ to ${\cal C}^1(\R)$, for any small $\eps > 0$
there is $C > 0$ such that
$$
\| w_{\eps} \|_{L^{\infty}} = \| \hat{w}_{\eps} \|_{L^{\infty}}
\leq C \| \hat{w}_1 + \hat{w}_2 + \hat{\varphi}_{\eps} \|_{H^2} \leq C \eps^{2/3},
$$
which completes the proof of bound (\ref{bound-second-excited-state}) for
any $(a,\zeta)$ satisfying a priori bounds (\ref{bounds-on-a}). It remains
to show that bounds (\ref{bounds-on-a}) are satisfied by solutions of the
Lyapunov--Schmidt bifurcation equation (\ref{LSreductions}).

{\bf Step 6: Lyapunov--Schmidt bifurcation equation.} To consider solutions of
the Lyapunov--Schmidt reduction equation, we rewrite (\ref{LSreductions}) in the form
$$
{\cal F}_{\eps} \equiv {\cal F}_{\eps}^{(1)} + {\cal F}_{\eps}^{(2)},
$$
where
\begin{eqnarray*}
{\cal F}_{\eps}^{(1)} & = & \langle \hat{\psi}^+_0, \hat{H}_{\eps} \rangle_{L^2}, \\
{\cal F}_{\eps}^{(2)} & = & \langle \tilde{\psi}^+_{\eps},\left( \hat{G}_{\eps} + {\cal H}_{\eps} +
{\cal N}_{\eps}(\hat{\varphi}_{\eps}) \right) \rangle_{L^2} +
\langle \tilde{\psi}^+_{\eps} - \hat{\psi}^+_0, \hat{H}_{\eps} \rangle_{L^2}.
\end{eqnarray*}
We will show that there exists a simple root of ${\cal F}_{\eps}^{(1)}$
in $a > 0$, which satisfies the asymptotic expansion (\ref{bound-a-asymptotic}) and
that this root persists with respect to the perturbations in ${\cal F}_{\eps}^{(2)}$.
If $a$ satisfies the asymptotic expansion (\ref{bound-a-asymptotic}),
then $a = {\cal O}(\eps |\log(\eps)|)$ and $e^{-2\zeta} = {\cal O}(\eps |\log(\eps)|^{1/2})$
so that a priori bounds (\ref{bounds-on-a}) are satisfied.

For convenience, we recall (\ref{eigenfunction-1}) and write
\begin{eqnarray}
\label{remainder-R}
{\cal R}_{\eps} \equiv \frac{2 \sqrt{2}}{\sqrt{3}} {\cal F}_{\eps}^{(1)} =
\langle ({\rm sech}^2(z_+) + {\rm sech}^2(z_-) + {\cal O}_{L^{\infty}}(e^{-2\zeta}),
\hat{H}_{\eps} \rangle_{L^2}.
\end{eqnarray}
In what follows, we compute the leading order of ${\cal R}$ and
account the error of the size ${\cal O}_{L^{\infty}}(e^{-2\zeta})$
in the end of computations.
From explicit definition of $\hat{H}_{\eps}$, the leading-order part of ${\cal R}_{\eps}$
is written in the form
\begin{eqnarray*}
{\cal R}_{\eps}^{(1)} & = & \langle ({\rm sech}^2(z_+) + {\rm sech}^2(z_-),
\hat{H}_{\eps} \rangle_{L^2} \\
& = & \int_{\R} \eta_{\eps}(x) (\eta_{\eps}^2(x) - 1) \tanh(z_+) \tanh(z_-) \left(
{\rm sech}^2(z_+) + {\rm sech}^2(z_-) \right)^2  dz \\
& \phantom{t} & + \sqrt{2} \eps \int_{\mathbb{R}} \eta_{\eps}'(x)
\left( \tanh(z_+) {\rm sech}^2(z_-) +  \tanh(z_-) {\rm sech}^2(z_+) \right) \left(
{\rm sech}^2(z_+) + {\rm sech}^2(z_-) \right)dz \\
& \phantom{t} & + \int_{\mathbb{R}} \eta_{\eps}(x) {\rm sech}^2(z_+) {\rm sech}^2(z_-) \left( 1 - \eta_{\eps}^2(x)
\tanh(z_+) \tanh(z_-) \right) \left(
{\rm sech}^2(z_+) + {\rm sech}^2(z_-) \right) dz.
\end{eqnarray*}
After the change of variables $u = z - \zeta = z_- = z_+ - 2\zeta$
and the use of symmetry on $z \in \mathbb{R}$, the first and second terms in
${\cal R}_{\eps}$ give
\begin{eqnarray*}
I_1 + I_2 & = & 2 \int_{-\zeta}^{\infty} \eta_{\eps}(x) (\eta_{\eps}^2(x) - 1)
\tanh(u) \tanh(u+2\zeta) \left(
{\rm sech}^2(u) + {\rm sech}^2(u + 2\zeta) \right)^2  du \\
& \phantom{t} & + 2 \sqrt{2} \eps \int_{-\zeta}^{\infty}  \eta_{\eps}'(x)
\left( \tanh(u) {\rm sech}^2(u + 2 \zeta) +  \tanh(u + 2\zeta) {\rm sech}^2(u) \right) \\
& \phantom{t} & \phantom{texttexttexttexttexttexttexttext} \times \left(
{\rm sech}^2(u) + {\rm sech}^2(u + 2\zeta) \right)du \\
& = & \frac{3 \sqrt{2} \eps}{2} \int_{-\zeta}^{\infty}
(1 + \eta^2_{\eps}(x)) \eta_{\eps}'(x)
{\rm sech}^4(u) \left(1 + {\cal O}_{L^{\infty}}(e^{-2\zeta}) \right) du
\end{eqnarray*}
where $x = \sqrt{2} \eps (u + \zeta)$. Thanks to the exponential decay
of ${\rm sech}^4(u)$ and property (P3), we have
\begin{equation}
\label{I1-I2}
I_1 + I_2 =  \frac{3 \sqrt{2} \eps}{2} \int_{-\zeta}^{\zeta}
(1 + \eta^2_{\eps}(x)) \eta_{\eps}'(x)
{\rm sech}^4(u) \left(1 + {\cal O}_{L^{\infty}}(e^{-2\zeta}) \right) du +
{\cal O}(\eps^{2/3} e^{-4 \zeta}).
\end{equation}
On the other hand, thanks to property (P2) for $\zeta \leq \beta \eps^{-1/3}$
for any $\beta \in (0,1)$, we have
$$
\eta_{\eps}(x) = 1 + {\cal O}_{L^{\infty}}(\eps^{4/3}), \quad \eta_{\eps}'(x) = -x (1 + {\cal O}_{L^{\infty}}(\eps^{4/3})), \quad
\forall x \in [0,2 \sqrt{2} \eps \zeta].
$$
As a result, we obtain
\begin{eqnarray*}
I_1 + I_2 & = & -6 \eps^2 \int_{-\zeta}^{\zeta} (\zeta + u) {\rm sech}^4(u) \left( 1 + {\cal O}_{L^{\infty}}(\eps^{4/3},e^{-2\zeta}) \right)  du +
{\cal O}(\eps^{2/3} e^{-4 \zeta}) \\
& = & -4 \sqrt{2} \eps a \left( 1 + {\cal
O}(\eps^{4/3},e^{-2\zeta}) \right) +
{\cal O}(\eps^{2/3} e^{-4 \zeta}).
\end{eqnarray*}
Performing similar computations for the third term in
${\cal R}_{\eps}$ gives
\begin{eqnarray*}
I_3 & = & 2 \int_{-\zeta}^{\infty} \eta_{\eps}(x)
{\rm sech}^4(u) {\rm sech}^2(u+2\zeta) \left( 1 - \eta_{\eps}^2(x)
\tanh(u) \right) \left(1 + {\cal O}_{L^{\infty}}(e^{-2\zeta}) \right) du \\
& = & 2^8 e^{-4 \zeta} \int_{-\zeta}^{\zeta}
\frac{e^{-8 u}}{(1 + e^{-2 u})^5} \left(1 + {\cal O}_{L^{\infty}}(\eps^{4/3},e^{-2\zeta}) \right) du
+ {\cal O}(e^{-6 \zeta})\\
& = & 32 e^{-4 \zeta} \left( 1 + {\cal
O}(\eps^{4/3},e^{-2\zeta}) \right) + {\cal O}(e^{-6 \zeta}).
\end{eqnarray*}
Recalling now (\ref{remainder-R}), we have thus obtained that
$$
{\cal R}_{\eps} = -4 \sqrt{2} \eps a \left( 1 + {\cal
O}(\eps^{2/3},e^{-2\zeta}) \right) + 32 e^{-2\sqrt{2} a \eps^{-1}} \left( 1 + {\cal
O}(\eps^{4/3},e^{-2\zeta}) \right).
$$
Analyzing similarly the error coming from the other term ${\cal F}_{\eps}^{(2)}$
in the Lyapunov--Schmidt reduction equation (\ref{LSreductions}), we rewrite
this equation in the form
\begin{equation}
\label{nonlinear-equation}
\frac{2 \sqrt{2}}{\sqrt{3}} {\cal F}_{\eps} = -4 \sqrt{2}
\eps a \left( 1 + {\cal O}(\eps^{2/3},e^{-2\zeta}) \right) + 32
e^{-2\sqrt{2} a \eps^{-1}} \left( 1 + {\cal
O}(\eps^{2/3},e^{-2\zeta}) \right) = 0.
\end{equation}
Taking a natural logarithm of ${\cal F}_{\eps} = 0$, we obtain
$$
2 \sqrt{2} a \eps^{-1} + \log(a) = -\log(\eps) + \frac{5}{2}
\log(2) + {\cal O}(\eps^{2/3},e^{-2\zeta}).
$$
Let $a = -\frac{1}{\sqrt{2}} \eps \log(\eps) U$ and rewrite the
problem for $U$:
\begin{equation}
\label{equation-U}
U - \frac{\log(U)}{2 \log(\eps)} =  1 + \frac{\log|\log(\eps)|}{2
\log(\eps)} - \frac{3\log(2)}{2 \log(\eps)} \left( 1 + {\cal
O}(\eps^{2/3},e^{-2\zeta}) \right).
\end{equation}
The remainder term is continuous with respect to $\eps$ for small $\eps > 0$.
There exists a root of (\ref{equation-U}) at $U = 1$ for $\eps = 0$.
By the Implicit Function Theorem applied to equation (\ref{equation-U})
for small $\eps > 0$, there exists a unique root $U(\eps)$ such that $U(\eps)$ is
continuous in $\eps > 0$ and $\lim_{\eps \downarrow 0} U(\eps) =
1$. To estimate the remainder term, one can further decompose
$$
U = 1 + \frac{\log|\log(\eps)|}{2 \log(\eps)} (1 + V)
$$
and rewrite the problem for $V$:
\begin{equation}
\label{equation-V}
V - \frac{\log\left( 1 + \frac{\log|\log(\eps)|}{2 \log(\eps)} (1
+ V) \right)}{\log|\log(\eps)|} = - \frac{3 \log(2)}{\log
|\log(\eps)|} \left( 1 + {\cal O}(\eps^{2/3},e^{-2\zeta}) \right).
\end{equation}
Again, there is a root of (\ref{equation-V}) at $V = 0$ for $\eps = 0$.
By the Implicit Function Theorem applied to equation (\ref{equation-V})
for small $\eps > 0$, there exists a unique root $V(\eps)$ such that $V(\eps)$ is
continuous in $\eps > 0$ and $\lim_{\eps \downarrow 0} V(\eps) =
0$. As a result, for small $\eps > 0$ there is a root of the nonlinear equation
(\ref{nonlinear-equation}), which admits the asymptotic expansion
(\ref{bound-a-asymptotic}).

{\bf Step 7: Properties (\ref{properties-uu}).} The uniform bound
(\ref{bound-second-excited-state}) has again the order of ${\cal
O}(\eps^{2/3})$. Using the same analysis as in Step 6 of
the proof of Theorem \ref{proposition-excited-state}, we prove that
$u_{\eps}(x)$ is strictly positive for any $|x| \geq 1$.
Therefore, there exist only two zeros of $u_{\eps}(x)$ on $\R$
and the two zeros $x = \pm x_0$ are located near $x = \pm a$
(Remark \ref{remark-zeros}). Additionally, $u_{\eps} \in {\cal C}^1(\R)$
and the bootstrapping arguments give $u_{\eps} \in {\cal C}^{\infty}(\R)$.
Combining all together, $u_{\eps}(x)$ constructed above
is the second excited state of the stationary equation (\ref{stationaryGPexc}) that satisfies
property (\ref{properties-uu}).

\section{Construction of the $m$-th excited state with $m \geq 2$}

The $m$-th excited state is constructed similarly to the proof of Theorem \ref{proposition-second-excited-state}.
The relevant decomposition is a product of $m$ dark solitons and the ground state in the form
$$
u_{\eps}(x) = \eta_{\eps}(x) \prod_{j = 1}^m \tanh\left(\frac{x - a_j}{\sqrt{2} \eps}\right)
+ w_{\eps}(x),
$$
where parameters $\{ a_j \}_{j = 1}^m$ are to be found from $m$ constraints on the
fixed-point problem for the remainder term $w_{\eps}(x)$. Assuming that
all $a_j$ are distinct and distributed according to the a priori bounds
\begin{equation}
\label{bounds-on-aa} \left\{ \begin{array}{l} \exists \beta \in (0,1) : \quad
a_j \leq \sqrt{2} \beta \eps^{2/3}, \quad j \in \{1,2,...,m\} \\
\exists C > 0 : \quad e^{-\sqrt{2} (a_{j+1} - a_j) \eps^{-1}} \leq C \eps^2 |\log(\eps)|, \quad
j \in \{1,2,...,m-1\}, \end{array} \right.
\end{equation}
the relevant potential of the Schr\"{o}dinger operator
$$
\hat{L}_0(a_1,...,a_m) = -\frac{1}{2} \partial_z^2 + 2 - 3 \sum_{j = 1}^m {\rm sech}^2(z - z_j), \quad
z_j = \frac{a_j}{\sqrt{2} \eps}
$$
has $m$ wells and supports $m$ eigenvalues in the neighborhood of $0$.
The $m$ constraints follow from $m$ projections to
the corresponding eigenfunctions for the $m$ smallest eigenvalues. Although the computations
of these reductions are long and cumbersome, these computations
are expected to recover the same leading order as the Euler--Lagrange equations obtained by Coles {\em et al.} \cite{CKP},
\begin{equation}
\label{Toda-lattice}
4 \sqrt{2} \eps a_j + 32 \left( e^{-\sqrt{2}(a_{j+1}-a_j) \eps^{-1}} -
e^{-\sqrt{2}(a_{j}-a_{j-1}) \eps^{-1}} \right) = 0, \quad j \in \{1,2,...,m\},
\end{equation}
where only pairwise interactions contribute to the leading order.
Asymptotic expansions of solutions of these equations are constructed in \cite{CKP}
and compared to the numerical approximations for $m = 2$ and $m = 3$.

Spectral stability of the excited states in the limit $\eps \to 0$ is also a physically
important and mathematically interesting problem. Variational and numerical approximations
in \cite{CKP} suggest that the purely discrete spectrum of the spectral stability problem
associated with the $m$-th excited state has a countable set of eigenvalues, which are close
to eigenvalues associated with the ground state, and $m$ additional pairs of eigenvalues.
The $m$ additional pairs are related to the Jacobian of the reductions equations (\ref{Toda-lattice}):
one pair remains bounded as $\eps \to 0$ and $(m-1)$ pairs grow like $\log(\eps)$
as $\eps \to 0$. Unfortunately, the rigorous studies
of the asymptotic properties of eigenvalues are difficult even for the
linearization of the ground state \cite{GalPel1}. Therefore, the characterization of
asymptotic properties of eigenvalues associated with the excited states will remain an
open problem for further studies.

{\bf Acknowledgement.} The author is thankful to Cl\'ement Gallo for careful reading of the manuscript
and critical remarks. A part of this work was supported by the NSERC grant.

\end{document}